\documentclass[aps,prb,reprint,showpacs,superscriptaddress]{revtex4-1}

\usepackage{graphicx}
\usepackage{bm}
\usepackage{dcolumn}
\usepackage{amssymb}
\usepackage{amsmath}
\usepackage{amsfonts}
\usepackage{color}
\usepackage{epsfig}
\usepackage{hyperref}
\usepackage{verbatim} 

\def \da{^{\dagger}}

\def \d{\mathrm{d}}

\def \H0{H_0}
\def \HL{H_{\mathrm{L}}}
\def \HR{H_{\mathrm{R}}}
\def \HD{H_{\mathrm{D}}}

\def \HTL{H_{\mathrm{TL}}}
\def \HTR{H_{\mathrm{TR}}}

\def \GRLin{\Gamma_{\mathrm{RL}}}

\begin{document}

\title{ Statistics of energy dissipation in a quantum dot operating in the 
cotunneling regime }

\author{Yehuda Dinaii}
\affiliation{Department of Condensed Matter Physics, The Weizmann Institute of
             Science, Rehovot 76100, Israel}

\author{Alexander Shnirman}
\affiliation{Institut f\"ur Theorie der Kondensierten Materie and DFG Center 
  for Functional Nanostructures (CFN), Karlsruhe Institute of Technology, 
  76128 Karlsruhe, Germany}
             
\author{Yuval Gefen}
\affiliation{Department of Condensed Matter Physics, The Weizmann Institute of
             Science,  Rehovot 76100, Israel}
\affiliation{Institut f\"ur Nanotechnologie, Karlsruhe Institute of
             Technology, 76021 Karlsruhe, Germany}

\date{\today}

\begin{abstract}

At Coulomb blockade valleys inelastic cotunneling processes generate 
particle-hole excitations in quantum dots (QDs), and lead to energy 
dissipation. We have analyzed the probability distribution function (PDF) of 
energy dissipated in a QD due to such processes during a given time interval.  
We obtained analytically the cumulant generating function, and extracted the 
average, variance and Fano factor. The latter diverges as $T^3/(eV)^2$ at bias 
$eV$ smaller than the temperature $T$, and reaches the value $3 eV / 5$ in the 
opposite limit. The PDF is further studied numerically.  As expected, Crooks 
fluctuation relation is not fulfilled by the PDF. Our results can be verified 
experimentally utilizing transport measurements of charge.

\end{abstract}


\maketitle

Thermal properties of nano-structures are of profound importance, inasmuch as 
they are manifestations of the dynamics of the particle zoo inside them. The 
latter includes electrons, phonons, photons, and other (quasi)particles, 
depending on the system and its surrounding environment. At the same time, 
understanding thermal characteristics and gaining the ability to manipulate 
them will facilitate higher control over nano-circuits, which is at the heart 
of technological advances. Importantly, it may push forward the effort towards 
finding sustainable energy resources.

As a consequence, recently there has been a growing interest in thermal 
aspects of nano-structures~\cite{Giazotto_2006, *Dresselhaus_2007, 
*Snyder_2008, *Benenti_2011}. For instance, thermoelectricity in semiconductor 
nano-structures is investigated in Ref.~[\onlinecite{Hsu_2004, *Lyeo_2004}].  
The validity of the Wiedemann-Franz law in several mesoscopic systems is 
studied in Refs.~[\onlinecite{Vavilov_2005, *Kubala_2008, *Balachandran_2012, 
*Lopez_2013}]. The temperature of nano-structures is analyzed in 
Refs.~[\onlinecite{Glatz_2010, *Heikkila_2009, *Laakso_2010a, *Laakso_2010b}].  
Verification of the recently discovered non-equilibrium fluctuation 
relations~\cite{Esposito_2009, *Campisi_2011} in the context of heat is 
reported in Refs.~[\onlinecite{Saira_2012, Utsumi_2014}]. Energy relaxation in 
a quantum dot (QD), which is a pillar in the study of nano-electronic systems, 
is investigated in Ref.~[\onlinecite{Chtchelkatchev_2009, 
*Chtchelkatchev_2013, *Chtchelkatchev_2013b}]. It was found there that half of 
the Joule-heating produced in transport is due to energy dissipation through 
the QD.  Importantly, there are physical phenomena which are not fully 
accessible by charge related measurements. As an example we note the recently 
observed neutral modes in the fractional quantum Hall regime~\cite{Bid_2010, 
*Takei_2011, *Dolev_2011, *Deviatov_2011, *Gross_2012, *Viola_2012, 
*Sabo_2013, *Inoue_2013b}, whose characterization may require 
thermometry~\cite{Venkatachalam_2012}.

\begin{figure}[tbp]
\begin{center}
\vspace{0cm}
\includegraphics[width=0.23\textwidth]{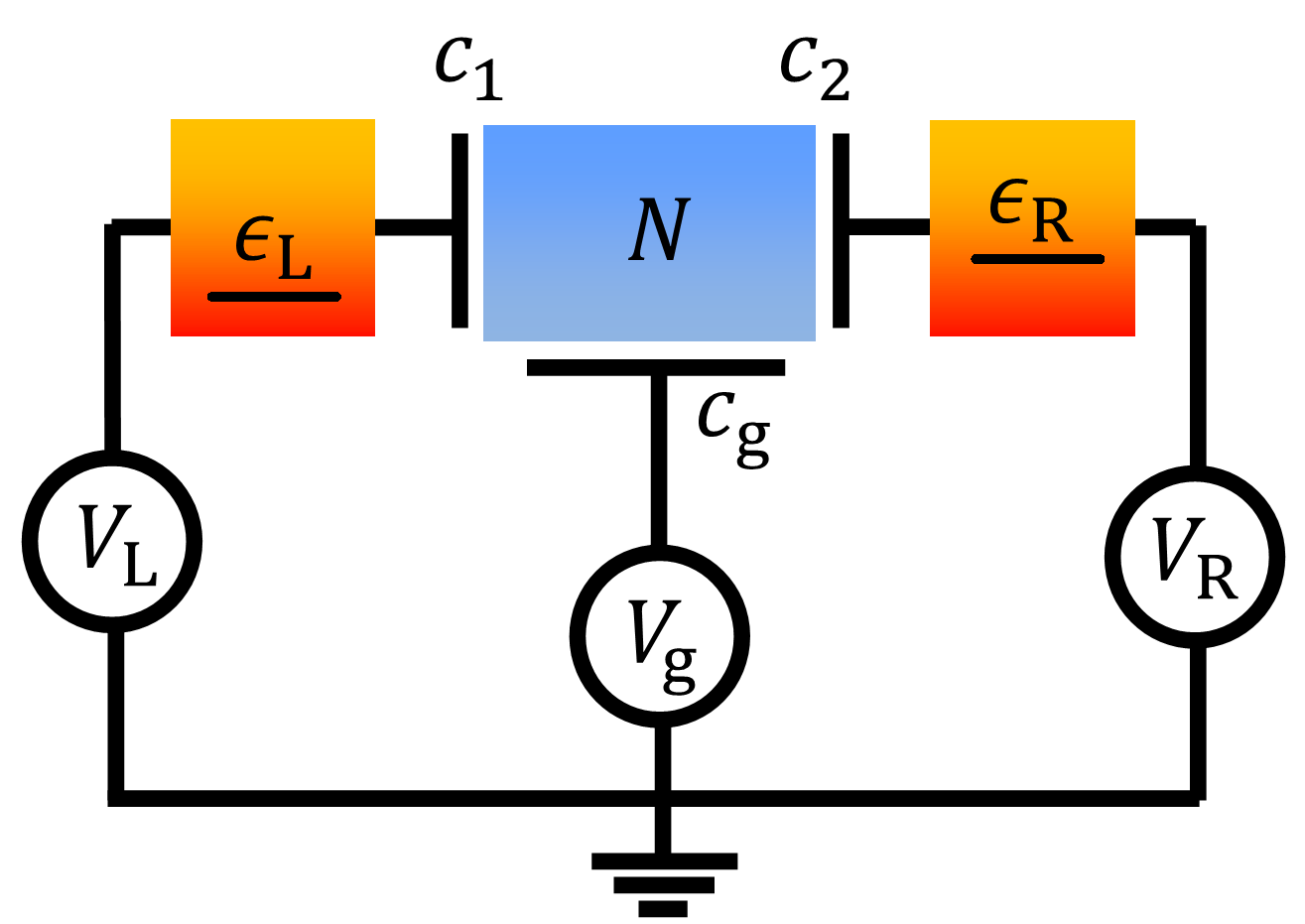}
\includegraphics[width=0.23\textwidth]{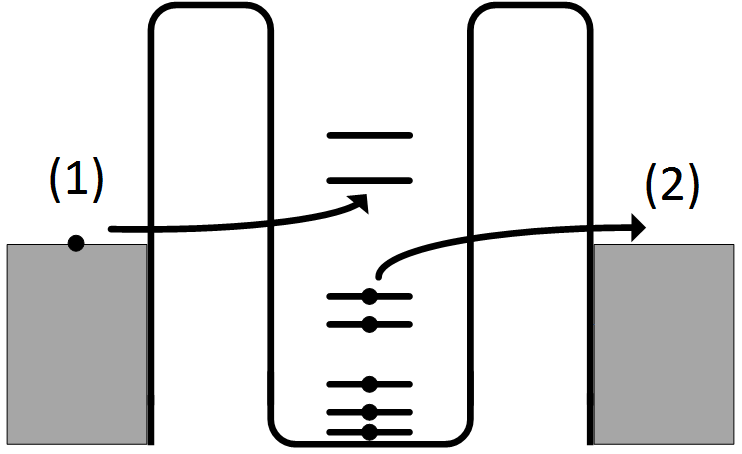}
\vspace{0cm}
\caption[] { \label{fig:setup}
Left: An equivalent circuit representing a quantum dot (QD) (the region 
bounded by the three capacitors $c_1$, $c_2$ and $c_\mathrm{g}$, marked by a 
blue rectangle) tunnel-coupled to two leads with potentials $V_{\mathrm{L}}$ 
and $V_{\mathrm{R}}$. The energy levels of the QD can be shifted by an 
additional capacitively-coupled gate $V_\mathrm{g}$. The two other orange 
rectangles denote energy filters. Right: Schematic illustration of a 
particle-like inelastic cotunneling process. The numbers denote the order of 
hopping. In the corresponding hole-like process the order is interchanged.
\vspace{-0.5cm}
}
\end{center}
\end{figure}

Here we study the statistical properties of energy dissipated in a 
QD~\cite{Alhassid_2000, *Aleiner_2002, *Pustilnik_2005} tuned to be in a 
Coulomb blockade valley. In this regime sequential tunneling processes are 
mostly suppressed, and cotunneling processes play a leading role in transport. 
Cotunneling is a many-body coherent process, where electrons are transferred 
from one lead to another via a virtual (classically forbidden) state in the 
QD~\cite{Averin_1989, *Averin_1990}. We are interested in the ``inelastic'' 
contribution, where a ``trace'' is left on the QD in the form of an 
electron-hole excited pair with energy $\Delta E$ (cf. Fig.~\ref{fig:setup}).  
Since the QD is practically always in contact with an environment, this energy 
is dissipated.  We focus on the regime where the time needed for equilibration 
of the QD constitutes the shortest time scale in the problem. The probability 
distribution function (PDF) $P(E, t)$ of the total energy dissipated in the 
QD, $E$, within a given time interval, $t$, possesses complete information on 
the statistics of energy dissipation in the QD.

\begin{figure}[tbp]
\begin{center}
\vspace{0cm}
\includegraphics[width=0.23\textwidth]{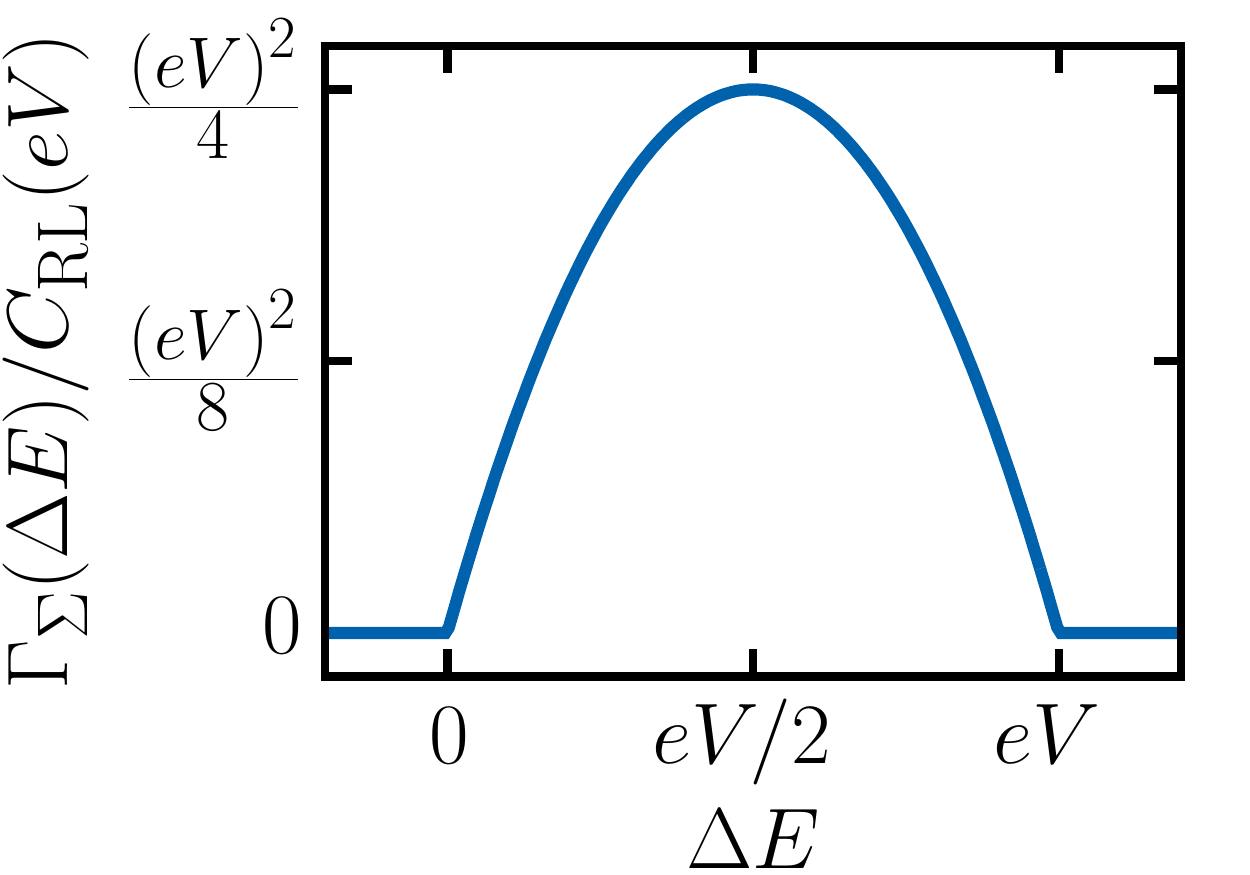}
\includegraphics[width=0.23\textwidth]{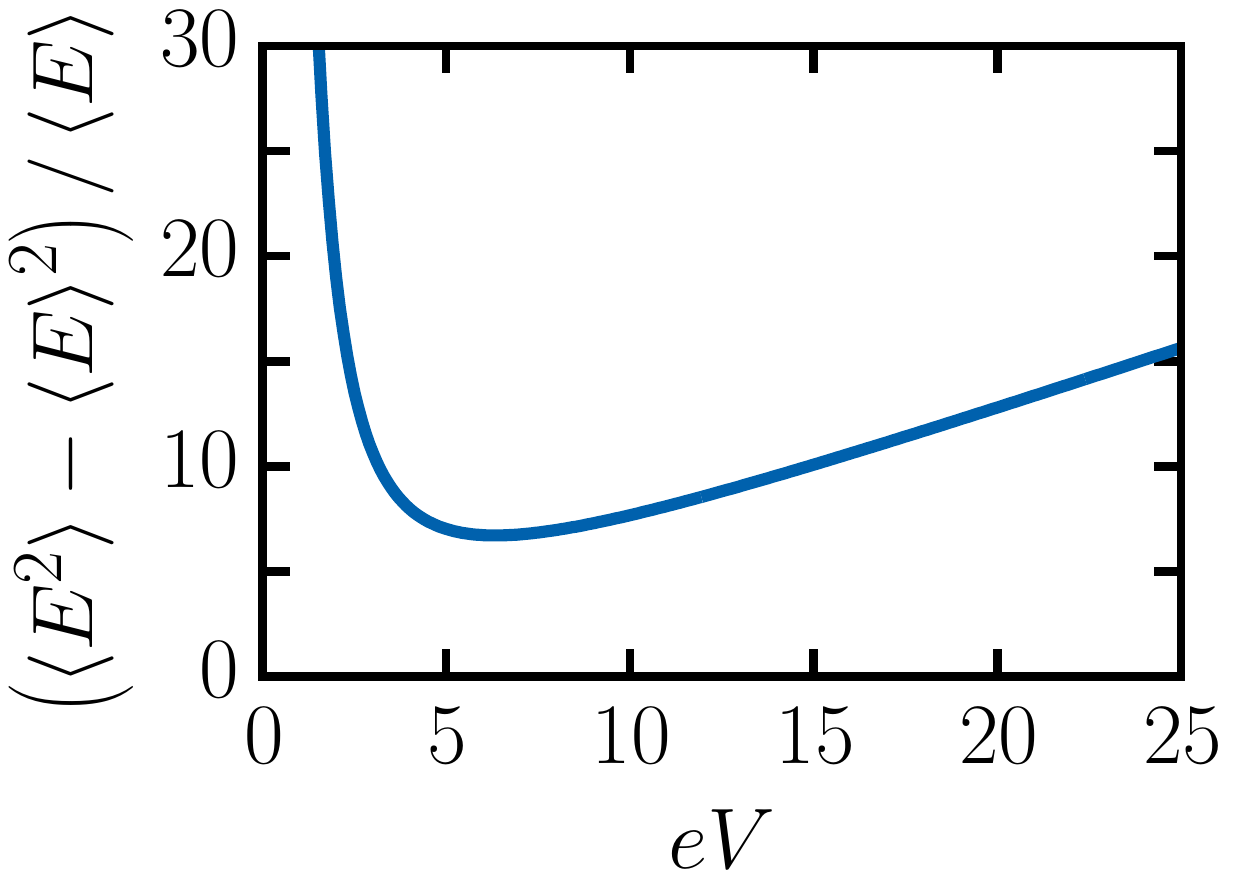}
\vspace{0cm}
\caption[] {Left: Total rate of energy dissipation in the QD,
$\Gamma_{\Sigma}(\Delta E)$, at temperature $T = 0$ (cf.  
Eq.~(\ref{eq:Gamma_RL_zero_T})). In this limit only inelastic cotunneling 
processes from left to right contribute to the energy dissipation in the QD, 
which is confined to the range $0<\Delta E<eV$. Right: Fano factor  (cf.  
Eq.~(\ref{eq:fano_sym})).
Here the temperature $T = 1$. At $eV \gg T$ the Fano factor $\sim 3eV/5$.  At 
$eV \rightarrow 0$ the divergence is a manifestation of the fact that on 
average no energy is dissipated in the QD, while fluctuations around this 
value are finite (cf. Eqs.~(\ref{eq:grp_sym})).
\label{fig:fano}
\vspace{-0.5cm}
}
\end{center}
\end{figure}

The main goal of our study is to tackle the PDF of energy dissipation in the 
context of virtual (classically forbidden) many-body states. Specifically, we
obtain the following: (i) An analytic result for the cumulant generating 
function of $P(E,t)$ (cf. Eq.~(\ref{eq:Gamma_sigma_in_tau_result})).  This 
function fully characterizes the statistics of energy dissipation in the QD, 
and can be utilized to obtain all the cumulants of the distribution. (ii) The 
PDF $P(E, t)$ in an integral form, which we study numerically. (iii) The first 
two cumulants of the PDF, average and variance (cf.  
Eq.~(\ref{eq:cumulants})), and the Fano factor (cf. Fig.~\ref{fig:fano}). (iv) 
We have analyzed our results in the context of non-equilibrium fluctuation 
relations, and have found that the PDF violates the Crooks relation. This is, 
in fact, expected, since the energy accounted for by the PDF is not the total 
work performed by the voltage source. (v) We propose an experimental method 
whereby statistics of energy can be acquired via charge-transport 
measurements.

The Hamiltonian of a QD tunnel-coupled to two leads (cf. Fig.~\ref{fig:setup}) 
is denoted by $H = H_{0} + V$. The unperturbed Hamiltonian $\H0 = \HL + \HR + 
\HD$ is the Hamiltonian of the three subsystems in the absence of tunneling, 
where $\HL = \sum_{k} \varepsilon_{k} c_{k}\da c_{k}$, $\HR = \sum_{q} 
\varepsilon_{q} c_{q}\da c_{q}$, $\HD = \sum_{n} \varepsilon_{n} c_{n}\da 
c_{n} + H_{N}$ are the Hamiltonians of the left-lead, right-lead, and the QD, 
respectively. $H_{N}$ denotes the interactions in the QD in the presence of 
$N$ electrons. The tunneling Hamiltonian is considered to be the perturbation.  
It is $V = \HTL + \HTR$, where $\HTL = \sum_{k,n} t_{kn} c_{k}\da c_{n} + 
\mathrm{H.C.}$ and $\HTR = \sum_{q,m} t_{qm} c_{q}\da c_{m} + \mathrm{H.C.}$ 
denote dot-left-lead tunneling and dot-right-lead tunneling, respectively.  

We employ the second-order version of Fermi's golden rule~\cite{*[{See e.g. }] 
[{, $\S 43$.}] Landau_1991_QM} to calculate the cotunneling rates of electrons 
from lead to lead~\cite{*[{See e.g. }] [{, $\S 10.3$.}] Bruus_2004} that 
deposit energy $\Delta E$ in the QD (which may be positive or negative) in the 
form of a particle-hole excitation (cf. Fig.~\ref{fig:setup}). For the 
transition rate per unit energy from the left-lead to the right-lead 
($\mathrm{L} \rightarrow \mathrm{R}$) we obtain
\begin{multline}
  \GRLin(\Delta E) = \frac{\gamma^{\mathrm{L}} \gamma^{\mathrm{R}}} {2\pi} 
  \int_{-\infty}^{+\infty}\!\!\!\!\!  \d\varepsilon_{k} 
  \int_{-\infty}^{+\infty} \!\!\!\!\!  \d \varepsilon_{n} 
  \int_{-\infty}^{+\infty}\!\!\!\!\!  \d\varepsilon_{m}\times\\
\times\int_{-\infty}^{+\infty}\d\varepsilon_{q} f(\varepsilon_{k}) \left[ 
  1 - f(\varepsilon_{n})\right]\delta(\varepsilon_{n}-\varepsilon_{m}-\Delta 
    E)\times\\
\times f(\varepsilon_{m}) \left[ 
  1 - f(\varepsilon_{q})\right]\delta\left(\varepsilon_{q} - 
    \varepsilon_{k}+\varepsilon_{n}-\varepsilon_{m} - eV \right)\times\\
    \times \left| \frac{1}{\varepsilon_{k} - \varepsilon_{n} + eV_{\mathrm{L}} 
  - \mu_N} + \frac{1}{\varepsilon_{m} - \varepsilon_{q} + \mu_{N-1} - 
  eV_{\mathrm{R}} } \right|^{2}\,.
\label{eq:inelastic_cotunneling_rate_2_Delta_E}
\end{multline}
Here $\mu_{N}\equiv e^{2} / 2c_{\Sigma} + e \left( eN + Q_{\mathrm{g}} \right) 
/ c_{\Sigma}$ is a charging energy associated with electron processes and 
$\mu_{N-1}$ is a charging energy associated with hole processes; $e$ is the 
charge of an electron, $c_{\Sigma} = c_1 + c_2 + c_{\mathrm{g}}$ the total 
capacitance of the QD to the leads and gate (cf.  Fig.~\ref{fig:setup}), 
$Q_{\mathrm{g}}$ the effective charge on the gate, and $e V \equiv e 
V_{\mathrm{L}} - e V_{\mathrm{R}} > 0$ the bias voltage.  It is assumed that 
the occupation of electronic states in each of the subsystems can be described 
by a Fermi function $f(\varepsilon) = (e^{\varepsilon/T}+1)^{-1}$ (i.e., fast 
relaxation time). Although the temperature in the leads and in the QD may 
differ~\cite{Heikkila_2009, Laakso_2010a, Laakso_2010b}, for simplicity, in 
what follows we assume that the temperature is uniform across the system. Our 
results are easily generalizable for the case of a higher steady state 
temperature in the QD. The constants $\gamma^{\mathrm{L(R)}} = 2\pi 
\rho_{\mathrm{L(R)}} \rho_{\mathrm{D}} \left| t_{kn(qm)} \right|^{2}$, where 
$\rho_{\mathrm{L(R)}}$ is the density of states in the left (right) lead, are 
assumed to be energy-independent. The energies $\varepsilon_{k}$, 
$\varepsilon_{n}$, $\varepsilon_{m}$, $\varepsilon_{q}$ correspond to levels 
in the left-lead, QD, QD, right-lead, respectively.

Similar expressions can be obtained for the rates of the other cotunneling 
processes, namely from the right-lead to the left-lead, from the left-lead to 
itself, and from the right-lead to itself. The total rate is given by 
$\Gamma_{\Sigma}(\Delta E) = \sum_{s,s' =
\mathrm{L}, \mathrm{R}} \Gamma_{ss'}(\Delta E)$, where
$\Gamma_{\mathrm{RL}}(\Delta E) = \tilde{\Gamma}_{\mathrm{RL}}(\Delta E, eV)$, 
$\Gamma_{\mathrm{LR}}(\Delta E) = \tilde{\Gamma}_{\mathrm{LR}}(\Delta E, 
-eV)$, $\Gamma_{\mathrm{LL}}(\Delta E) = \tilde{\Gamma}_{\mathrm{LL}}(\Delta 
E, 0)$, $\Gamma_{\mathrm{RR}}(\Delta E) = \tilde{\Gamma}_{\mathrm{RR}}(\Delta 
E, 0)$. The rates marked with a tilde are given by
\begin{subequations}\label{eq:rates}
  \begin{align}
  \tilde{\Gamma}_{ss'}(\Delta E,eV) & \equiv \int_{-\infty}^{\infty} 
  \mathrm{d} \varepsilon\, P_{\mathrm{eh}}(\varepsilon,eV-\Delta E) \times 
  \label{eq:Gamma_general} \\
  & \times\int_{-\infty}^{\infty} \mathrm{d} \varepsilon'\, P_{\mathrm{eh}} 
  (\varepsilon',\Delta E) P_{\mathrm{cot}}^{ss'} 
  (\varepsilon,\varepsilon',\Delta E) \,,
  \nonumber \\
  P_{\mathrm{eh}}(\varepsilon,\Delta E) & \equiv f(\varepsilon) \left[ 1 - 
    f(\varepsilon + \Delta E) \right] \,, \label{eq:pdf_of_eh}\\
  P_{\mathrm{cot}}^{ss'}(\varepsilon,\varepsilon',\Delta E) & \equiv 
  \gamma^{s}\gamma^{s'} (\mu_{N} - \mu_{N-1})^{2} / 2\pi \times  
  \label{eq:P_cot}\\
  & \times (\varepsilon' - \varepsilon + \mu_{N} - eV_{s'} + \Delta E)^{-2} 
  \times \nonumber \\
  & \times (\varepsilon' - \varepsilon + \mu_{N-1} - eV_{s'} + \Delta 
  E)^{-2}\,.\nonumber
 \end{align}
\end{subequations}
The quantity $P_{\mathrm{eh}}(\varepsilon,\Delta E)$ represents the 
probability for electron-hole excitations, and $P_{\mathrm{cot}}^{ss'} 
(\varepsilon, \varepsilon', \Delta E)$ has the meaning of a probability of a 
cotunneling process which leaves energy $\Delta E$ in the QD.

For temperatures and voltages small relative to the charging energy of the QD, 
this analysis can be further pursued analytically. We expand the integrands up 
to first order with respect to the kinetic energies over the charging 
energies, and evaluate the integrals. The result is
\begin{equation}
  \!\!\! \tilde{\Gamma}_{ss'}(\Delta E, eV) \simeq  C_{ss'}(eV) b(-\Delta E) 
  b(\Delta E - eV) \Delta E(eV - \Delta E)\,,
\label{eq:expansion}
\end{equation}
where $b(\varepsilon) = (e^{\varepsilon/T} - 1)^{-1}$ is the Bose function, 
and
\begin{multline}
  C_{ss'}(eV) \equiv \frac{\gamma^{s} \gamma^{s'}}{2\pi} \left( \frac{1}{
  \mu_{N-1} - eV_{s'}} - \frac{1}{\mu_{N}-eV_{s'}} \right)^{2} \times \\
  \times \left[1 - \left(\frac{1}{\mu_{N-1} - eV_{s'}} + \frac{1}{\mu_{N} -
  eV_{s'}}\right) eV \right]\,.
  \label{eq:const}
\end{multline}

In order to obtain some physical intuition, we look now at the limit of zero 
temperature.  Eqs.~(\ref{eq:rates}) readily show that in this limit all rates 
vanish besides $\Gamma_{\mathrm{RL}}(\Delta E)$, due to the presence of the 
Fermi functions.  Furthermore, $0<\Delta E<eV$. This is expected, since at 
zero temperature the only way the QD can be excited is when an energetic 
electron starts at the left-lead and passes to the right-lead while depositing 
some energy in the QD.  All other transitions are impossible, due to the 
filled Fermi seas in the left-lead, right-lead and QD.  
Eq.~(\ref{eq:expansion}) then yields at $T=0$
\begin{equation}
  \Gamma_{\Sigma}(\Delta E) \simeq
  \begin{cases}
    C_{\mathrm{RL}}(eV) \Delta E(eV-\Delta E)\,, & 0<\Delta E<eV \\
    0\,, & \mathrm{elsewhere}
  \end{cases} \, .
  \label{eq:Gamma_RL_zero_T}
\end{equation}
This is depicted in Fig.~\ref{fig:fano}.

We turn now to the calculation of $P(E,t)$, which denotes the PDF of the QD to 
absorb an excessive amount of energy $E$ during the time interval $t$ due to 
inelastic cotunneling processes. It is assumed that any amount of energy 
transferred to the QD due to a cotunneling electron immediately dissipates to 
the environment, namely that the relaxation time of the QD to an equilibrium 
state is the shortest time scale in the problem. $P(E,t)$ fulfills the 
following master equation,
\begin{multline}
  \frac{\partial P(E,t)}{\partial t} = - \Gamma_{\Sigma} P(E,t) \\
  +\int_{-\infty}^{\infty}\d(\Delta E) \Gamma_{\Sigma}(\Delta E) P(E - \Delta 
  E,t) \,,
  \label{eq:master_equation_b}
\end{multline}
where $\Gamma_{\Sigma} \equiv \int_{-\infty}^{\infty}\d(\Delta 
E)\Gamma_{\Sigma}(\Delta E)$ is the sum of rates of inelastic cotunneling at 
all energies. Taking the Fourier transform of Eq.~(\ref{eq:master_equation_b}) 
with respect to $E$ ($\tau$ will designate the variable conjugate to $E$) and 
solving the resulting differential equation one obtains
\begin{subequations}\label{eq:pdf}
  \begin{align}
  P(\tau,t) & = P(\tau,0) \exp \left\{ 2 \pi \left[ \Gamma_{\Sigma}(\tau) -
  \Gamma_{\Sigma}(\tau = 0) \right] t \right\} \,,
  \label{eq:pdf_a}\\
  P(E,t) & = \int_{-\infty}^{\infty} \mathrm{d}\tau \, P(\tau,t) e^{i E 
  \tau}\,.
  \label{eq:pdf_b}
 \end{align}
\end{subequations}
Here $\Gamma_{\Sigma}(\tau) = (2 \pi)^{-1} \int_{-\infty}^{\infty} \mathrm{d} 
(\Delta E) \, \Gamma_{\Sigma}(\Delta E) e^{-i \Delta E \tau}$. Normalization 
gives $\int_{-\infty}^{\infty} \mathrm{d}E\, P(E,t) = 2 \pi P(\tau = 0, t = 0) 
= \int_{-\infty}^{\infty} \mathrm{d} E\, P(E, t = 0)$, namely the PDF evolves 
in time such that the total probability is conserved, as it should. To 
facilitate the numerical evaluation of Eq.~(\ref{eq:pdf_b}), see below, we 
choose the initial condition $P(E, t = 0) = \exp \left( -E^{2} /2\sigma^{2} 
\right) / \sqrt{2 \pi \sigma^{2}}$. Physically it may reflect some initial 
uncertainty in the energy counter~\footnote{This initial
condition introduces an additional contribution to the cumulant generating 
function discussed below; namely $-\sigma^2 \tau^2 / 2$. This adds to the 
variance of the PDF a contribution equal to $\sigma^2$. We ignore this 
contribution in the analytical treatment below, since, in principle, $\sigma$ 
can be as small as one wishes}.

The cumulant generating function is given by $\ln \left\langle e^{i E \tau} 
\right\rangle = $ $2\pi \left[ \Gamma_{\Sigma}(-\tau) \right.$
$\left. -\Gamma_{\Sigma}(\tau = 0) \right] t$. Consequently, the $n^{
\mathrm{th} }$-cumulant is given by $2 \pi t \cdot  i^{n} \left.  
\partial_{\tau}^{n} \Gamma_{\Sigma}(\tau) \right|_{\tau = 0}$ where
\begin{multline}
  \!\!\!\!\!\! \Gamma_{\!\Sigma}( \tau \! ) \!\! = \!\! \frac{\pi T^{3} \!
  \left[ C_{\mathrm{LL}}(0) \! + \! C_{\mathrm{RR}}(0) \right]}{\sinh^{3}
  \left(\pi T \tau \right)} \! \left[\pi T \tau \! \cosh(\pi T \tau) \! - \!
  \sinh(\pi T\tau)\right] \\
  + \frac{\pi^2 T^{3} \left[ C_{\mathrm{RL}}(eV) e^{\frac{eV}{2}
  \left(\frac{1}{T} - i \tau \right)} + C_{\mathrm{LR}}(-eV) e^{-\frac{eV}{2} 
  \left(\frac{1}{T} - i \tau \right)} \right]}{\sinh\left( \frac{eV}{2T} 
  \right)\sinh^{3}\left( \pi T \tau \right)} \times \\
  \times\left[\sin \left(\frac{eV\tau}{2} \right)\cosh(\pi T\tau)
  - \frac{eV}{2\pi T}\cos \left(\frac{eV\tau}{2} \right)\sinh(\pi T \tau) 
  \right] \,.
  \label{eq:Gamma_sigma_in_tau_result}
\end{multline}
This function, which provides complete information on the statistics of energy 
dissipation in the QD upon differentiation, is the central result of our 
manuscript. As a consistency check we obtain the standard inelastic charge 
current~\cite{Averin_1989, *Averin_1990} from these results, which, for 
$C_{\mathrm{RL}}(eV) \simeq C_{\mathrm{LR}}(-eV)$, reads $I = 2 \pi e 
[\Gamma_{\mathrm{RL}} (\tau = 0) - \Gamma_{\mathrm{LR}} (\tau = 0) ] \propto 
eV [ \left( 2\pi T \right)^{2} + (eV)^{2} ]$.

The first two cumulants of $P(E, t)$ --- the mean value and the variance --- 
are given by
\begin{subequations}
\label{eq:cumulants}
\begin{align}
  \frac{\left\langle E \right\rangle}{t} & = 
  \frac{C_{\mathrm{RL}}(eV)e^{eV/2T} - 
  C_{\mathrm{LR}}(-eV)e^{-eV/2T}}{24\sinh(eV/2T)} \times \nonumber \\
  & \times \left( eV \right)^{2} \left[ \left( eV \right)^{2} + \left(2\pi T
  \right)^{2} \right] \,,
  \label{eq:k1}\\
  \frac{\left\langle E^{2} \right\rangle - \left\langle E 
  \right\rangle^{2}}{t} & = \frac{1}{30} [C_{\mathrm{LL}}(0) + 
    C_{\mathrm{RR}}(0)] (2 \pi T)^4 T \nonumber \\
  & + \frac{C_{\mathrm{RL}}(eV) e^{eV / 2T} +  C_{\mathrm{LR}}(-eV) 
  e^{-eV/2T}}{120 \sinh(eV / 2T)} eV \times \nonumber \\
  & \times \left[ \left( eV \right)^{2} + \left( 2 \pi T \right)^{2}
  \right] \left[ 3 \left( eV \right)^{2} + 2 \left( 2 \pi T \right)^{2}
  \right] \, ,
\label{eq:k2}
\end{align}
\end{subequations}
It is noted that $\left\langle E\right\rangle / t = IV/2$ (cf.  
Ref.~\onlinecite{Chtchelkatchev_2013}). Similarly, it is possible to evaluate 
higher-order cumulants of $P(E,t)$. 

In the symmetric case where $\gamma^{\mathrm{L}} = \gamma^{\mathrm{R}} \equiv 
\gamma$, and for values of $e V_{\mathrm{L}}$ and $e V_{\mathrm{R}}$ which are 
small relative to the charging energies, one has $C_{ss'}(eV) \simeq 
\left(\mu_{N - 1}^{-1} - \mu_{N}^{-1} \right)^{2} \gamma^{2} / 2 \pi \equiv 
C$. It follows that
\begin{subequations} \label{eq:grp_sym}
\begin{align}
  \frac{\left\langle E \right\rangle}{t} = \frac{C}{12} \left(eV\right)^{2} & 
  \left[
  \left( eV \right)^{2} + \left(2\pi T\right)^{2} \right] \,,
  \label{eq:k1_sym}\\
  \frac{\left\langle E^{2} \right\rangle - \left\langle E 
  \right\rangle^{2}}{t} = \frac{C}{60} & \left\{ (2 \pi T)^4 4 T + \coth 
    \left( \frac{eV}{2T} \right) eV \right. \times \nonumber \\
  \times \left[ \left(eV \right)^{2} + \left(2\pi T\right)^{2}\right] & \left.  
  \left[3 \left(eV \right)^{2} + 2\left(2\pi T\right)^{2} \right] \right\} \,,
  \label{eq:k2_sym} \\
  \frac{\left\langle E^{2}\right\rangle - \left\langle E \right\rangle^{2}}{ 
  \left\langle E \right\rangle} & =  \coth \left(\frac{eV}{2T} \right)    
  \frac{3 \left( eV \right)^{2} + 2\left(2 \pi T\right)^{2}}{5eV} \nonumber \\
  & + \frac{4T(2 \pi T)^{4}}{5 \left( eV \right)^{2}\left[(2 \pi T)^{2} +
  (eV)^{2} \right]} \,.
  \label{eq:fano_sym}
\end{align}
\end{subequations}
The information on the average and variance is encapsulated in the Fano 
factor, which is the ratio between them; it is shown in Fig.~\ref{fig:fano}. 

In the high bias regime, $eV \gg T$, one observes the following. The average 
$\left\langle E \right\rangle / t \propto (eV)^4$, implying that the QD is 
more probable to absorb energy than to emit energy. The fluctuations (i.e., 
standard deviation) $\propto (eV)^{5/2}$. The Fano factor in this limit 
$\simeq 3eV/5$, expressing a corresponding ``effective energy charge''.
  
The results in the linear response regime, $eV \ll T$, are quite different.  
The average $\left\langle E \right\rangle / t \propto (eV)^2 T^2$, and the 
fluctuations $\propto T^{5/2}$. This is reflected in the divergence of the 
Fano factor, which in this limit $\simeq 32 \pi^2 T^3 / 5 (eV)^2$, see 
Fig.~\ref{fig:fano}(b).

The Crooks fluctuation relation is not fulfilled by $P(E,t)$. In the present 
context the Crooks relation reads $P(E,t) = P(-E,t) e^{E/T}$, which upon 
Fourier transform yields $P(\tau, t) = P(-\tau-i/T, t)$. The latter relation 
is generically violated by $P(\tau,t)$ given in Eq.~(\ref{eq:pdf_a}).  This 
can be understood by recalling that the Crooks relation applies to the total 
energy (work) gained by a system, while here $E$ denotes only the energy 
gained by the QD (and not the energy dissipated in the left and right leads).  
As a consequence of the symmetry of the problem, $P(E,t)$ is unchanged with 
respect to a simultaneous interchange of $V_{\mathrm{L}} \rightleftharpoons 
V_{\mathrm{R}}$ and $\gamma^{\mathrm{L}} \rightleftharpoons 
\gamma^{\mathrm{R}}$.

\begin{figure}[tbp]
\begin{center}
\vspace{0cm}
\includegraphics[width=0.48\textwidth]{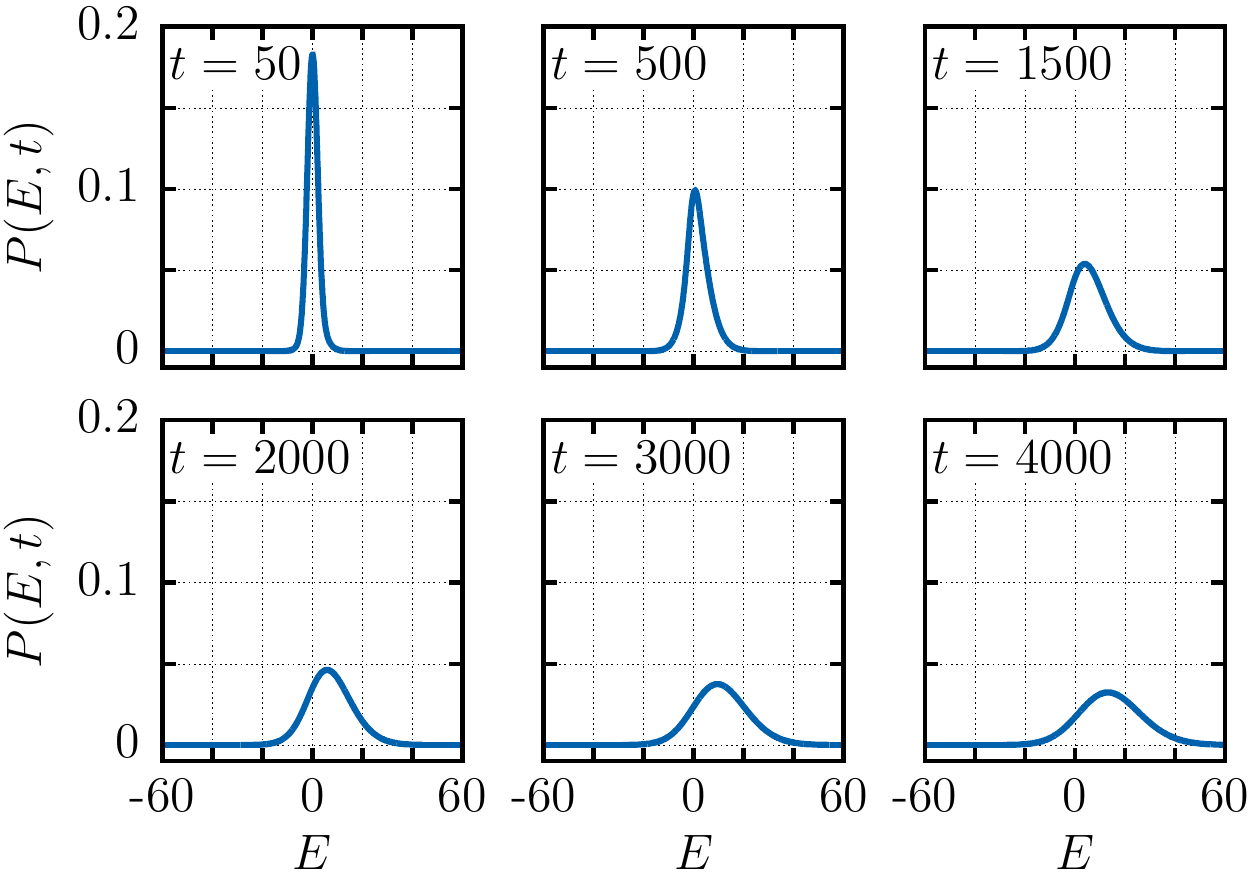}
\vspace{-0.5cm}
\caption[] {The time evolution of the probability distribution of energy
dissipated in the QD, $P(E,t)$ (cf. Eqs.~(\ref{eq:pdf},
\ref{eq:Gamma_sigma_in_tau_result})). The time intervals are indicated in the 
panels. $P(E,t)$ is obtained by numerical integration with $T=1$, $eV=3$, $C = 
10^{-4}$, and $\sigma = 2$.  The typical time scale associated with the 
evolution of $P(E,t)$ is given by $\Gamma_{\Sigma}^{-1}(\tau = 0) \simeq 
1570$.
\label{fig:time_evolution}
\vspace{-0.5cm}
}
\end{center}
\end{figure}

It is possible to evaluate $P(E,t)$ by performing the Fourier transform in 
Eq.~(\ref{eq:pdf_b}) numerically. The evolution of $P(E,t)$ for a case where 
$eV>T$ is shown in Fig.~\ref{fig:time_evolution}.  $P(E,t)$ is seen to 
propagate and widen, where the typical time scale of its evolution is given by 
$\Gamma_{\Sigma}^{-1}(\tau = 
0)$.  

\emph{Experimental considerations} --- One route to measure $P \left( E, t 
\right)$ is with sensitive thermometry~\cite{Giazotto_2006}. However, issues 
concerning ``back-action'' due to the measurement device may then 
arise~\cite{Utsumi_2014}. In what follows we propose another method, which is 
based on a transport measurement of charge.  We first conceive ideal energy 
filters deployed in the left and right leads (see Fig.~\ref{fig:setup}). These 
filters will allow only electrons with certain energies, say 
$\epsilon_{\mathrm{L}}$ and $\epsilon_{\mathrm{R}}$, to pass through. We 
define the rates of charge transfer at these energies, $\Gamma_{\mathrm{RL}} 
\left( \epsilon_{\mathrm{L}}, \epsilon_{\mathrm{R}} \right)$ and 
$\Gamma_{\mathrm{LR}} \left( \epsilon_{\mathrm{L}}, \epsilon_{\mathrm{R}} 
\right)$. A measurement of the current and noise, which are proportional to 
the difference and the sum of these rates, respectively, suffices for 
determining each of them separately~\footnote{The contribution of elastic 
cotunneling processes is negligible since the QD is metallic.}. Change of 
variables $\epsilon_{\mathrm{L}}, \epsilon_{\mathrm{R}} \rightarrow 
\epsilon_{\mathrm{L}} + \epsilon_{\mathrm{R}}, \pm (\epsilon_{\mathrm{R}} - 
\epsilon_{\mathrm{L}})$ and integration of $\Gamma_{\mathrm{RL}} \left( 
\epsilon_{\mathrm{L}}, \epsilon_{\mathrm{R}} \right)$ and 
$\Gamma_{\mathrm{LR}} \left( \epsilon_{\mathrm{L}}, \epsilon_{\mathrm{R}} 
\right)$ over $\epsilon_{\mathrm{L}} + \epsilon_{\mathrm{R}}$ then yield 
$\Gamma_{\mathrm{RL}} \left( \Delta E\right)$ and $\Gamma_{\mathrm{LR}} \left( 
\Delta E \right)$, respectively.
If the setup is symmetric, i.e., $\gamma^{\mathrm{L}} = \gamma^{\mathrm{R}}$, 
extraction of the two other rates, $\Gamma_{\mathrm{LL}} \left( \Delta E 
\right)$ and $\Gamma_{\mathrm{RR}} \left( \Delta E \right)$, is possible. At 
$e V = 0$ there is no net current, and the electric current noise is 
proportional to the sum of two equal rates, 
$\Gamma_{\mathrm{RL}}(\epsilon_{\mathrm{L}}, \epsilon_{\mathrm{R}})$ and 
$\Gamma_{\mathrm{LR}}(\epsilon_{\mathrm{L}},\epsilon_{\mathrm{R}})$. By taking 
$1/2$ of the measured noise we obtain each of those equal rates, as well as 
$\Gamma_{\mathrm{LL}}(\epsilon_{\mathrm{L}}, \epsilon_{\mathrm{L}}') = 
\Gamma_{\mathrm{RR}}(\epsilon_{\mathrm{R}}', \epsilon_{\mathrm{R}})$ with 
$\epsilon_{\mathrm{L}}' = \epsilon_{\mathrm{R}}$, $\epsilon_{\mathrm{R}}' = 
\epsilon_{\mathrm{L}}$.  At finite $eV$, the rates 
$\Gamma_{\mathrm{LL}}(\Delta E)$ and $\Gamma_{\mathrm{RR}}(\Delta E)$ remain 
unchanged. Note that restricting ourselves to zero temperature, the PDF is 
dominated now by a single rate ($\Gamma_{\mathrm{RL}}(\Delta E)$), and our 
analysis does not require the knowledge of $\Gamma_{\mathrm{LL}}(\Delta E)$ 
and $\Gamma_{\mathrm{RR}}(\Delta E)$. 

Two extra QDs tuned to resonances at energies $\epsilon_{\mathrm{L}}$ and 
$\epsilon_{\mathrm{R}}$ can be used to implement the energy filters. The 
resulting energy resolution will be of the order of the level width of the 
filters. We require that this width is determined by the coupling of the 
filter to the respective lead rather than to the central QD.

To further improve the energy resolution of the filters, one may introduce a 
junction with three entry/exit directions in between the QD and each of the 
filters. Each junction should be connected to the QD, to the nearby filter, 
and to an additional drain. By breaking the time reversal symmetry the 
junction can be tuned such that most backscattered electrons are drained out 
of the circuit through the additional drain and hence do not affect the 
measurement.

The results reported here constitute a step towards understanding energy 
characteristics of nanoscopic setups. Quantum dots, being a pillar of such 
systems, play an important role in such investigations. By studying a QD 
operating in the cotunneling regime, the energy characteristics of the QD in 
the ``deep'' quantum limit has been addressed directly.

\emph{To conclude}, we have analysed energy dissipation in a QD operating in 
the cotunneling regime, where energy is transferred to the QD in the form of 
particle-hole excitations. The QD is in contact with an environment, which 
supplies an equilibration mechanism to the excess energy deposited on the QD 
by the cotunneling electrons (this energy may also be negative). The time 
scale associated with the equilibration of the QD is assumed to be the 
shortest one in the system. We have analytically obtained the cumulant 
generating function, which supplies complete information on the statistics of 
energy dissipation in the QD. Specifically, the average, variance and Fano 
factor have been evaluated. We have further obtained numerically the 
corresponding PDF. The analysis of the results in the context of the recently 
discovered fluctuation relations underlines that fluctuation relations should 
be applied with caution. Our results are amenable to experimental verification 
with thermometry, or, with the more common transport measurement of charge.

Financial support by the German-Israel Foundation (GIF) and the Israel Science 
Foundation is acknowledged.

\bibliography{ref}

\end{document}